\documentclass[twocolumn,amsmath,amssymb,prl]{revtex4}
\usepackage{graphicx}
\usepackage{dcolumn}
\usepackage{bm}

\newcommand{\la}{\lesssim}
\newcommand{\ga}{\gtrsim}

\def\apj{Astrophys. J.~}

\def\mnras{Mon. Not. R. Astr. Soc.~}

\def\aap{Astr.\& Astrophys.~}

\def\beq{\begin{equation}}
\def\eeq{\end{equation}}

\begin{document}
\input{epsf}

\title{Observable Signatures of a Black Hole Ejected by Gravitational \\
Radiation Recoil in a Galaxy Merger}

\author{Abraham Loeb}

\affiliation{Astronomy Department, Harvard University, 60 Garden
Street, Cambridge, MA 02138, USA}

\begin{abstract}

According to recent general-relativistic simulations, the coalescence of
two spinning black holes (BHs) could lead to recoil speeds of the BH
remnant of up to thousands of km~s$^{-1}$ as a result of the anisotropic
emission of gravitational radiation. Such speeds would enable the merger
product to escape its host galaxy. Here we examine the circumstances
resulting from a gas-rich galaxy merger under which the ejected BH would
carry an accretion disk with it and be observable.  As the initial BH
binary emits gravitational radiation and its orbit tightens, a hole is
opened around it in the disk which delays the consumption of gas prior to
the eventual BH ejection. The punctured disk remains bound to the ejected
BH within the region where the gas orbital velocity is larger than the
ejection speed.  For a $\sim 10^7M_\odot$ BH the ejected disk has a
characteristic size of tens of thousands of Schwarzschild radii and an
accretion lifetime of $\sim 10^{7}$ years. During that time, the ejected BH
could traverse a considerable distance and appear as an off-center quasar
with a feedback trail along the path it left behind.  A small fraction of
all quasars could be associated with an escaping BH.

\end{abstract}


\maketitle

\paragraph{Introduction.}
The past few years witnessed a breakthrough in numerical relativity as
simulations were able to follow for the first time the final coalescence
phase of a binary black hole (BH) system due to the emission of
gravitational radiation
\cite{Centrella,Campanelli,Gonzales,Hermann,Koppitz,Schnitt,Tichy}.  While
the recoil speed of the remnant from binaries of non-spinning BHs is modest
($\la 200~{\rm km~s^{-1}}$), the coalescence of spinning BHs of nearly
equal masses could yield recoil speeds of up to thousands of km~s$^{-1}$
\cite{Campanelli,Schnitt,Tichy}.  The largest speeds are obtained for
special orientations of the spins relative to the orbital plane, and it is
unclear whether they occur in nature \cite{Bog}.  To find out if they do,
it is necessary to identify observational signatures of the ejected BHs.

Accretion of interstellar medium (ISM) gas by a fast-moving BH produces
only faint luminosities.  For a BH ejection speed $v_{\rm ej}$ which is
well above the sound speed of the ISM, the \citet{Bondi} accretion rate is
$\dot{M}_{\rm B}= 7\times 10^{-7} n_0 M_7^2v_8^{-3}~M_\odot~{\rm yr^{-1}}$,
where $M_7$ is the BH mass in units of $10^7M_\odot$, $v_8$ is the BH speed
in units of $10^8~{\rm cm~s^{-1}}=10^3~{\rm km~s^{-1}}$, and $n_{0}$ is the
ISM density in units of $1~{\rm cm^{-3}}$. Even if a thin accretion disk
forms around the ejected BH and the Bondi accretion rate is converted to
radiation with a high efficiency $\epsilon=0.1\epsilon_{-1}$, the resulting
luminosity $L=\epsilon\dot{M}_{\rm B}c^2=4\times
10^{39}\epsilon_{-1}n_{0}M_7^2v_8^{-3}~{\rm erg~s^{-1}}$, will be difficult
to detect at cosmological distances, except on the rare occasion when
the ejected BH passes through a dense molecular cloud.

The ejected BH could appear much brighter if it carries an accretion disk
with it. As long as the disk mass is much smaller than the BH mass and the
gas within the disk is orbiting at a speed far greater than the BH ejection
speed (which is possible since $v_{\rm ej}\ll c$), the gas will preserve
the adiabatic invariants of its orbit around the BH and follow the BH along
its ejection trajectory.

Next we characterize the properties of accretion disks that could be
carried by ejected BHs, and in the final section we discuss their
observational signatures.

\paragraph{Disk Parameters.}
Hydrodynamic simulations indicate that a gas-rich merger between a pair of
galaxies tends to drive their supermassive BHs together with some gas
towards the center of the merger product
\cite{Hernquist1,Hernquist2,Escala}. The latest simulations indicate that
the cold gas is not expelled by the feedback from early accretion episodes
when the two BHs are far apart, and so a gaseous envelope forms around the
coalescing BH binary.

The coalescence of the two BHs is driven at first by dynamical friction on
the gas and stars \cite{MM,Hernquist2,Escala}. Once the BH binary
separation shrinks to the regime where its orbital speed exceeds $\sim
10^{-2}c$, gravitational radiation alone could cause coalescence within a
Hubble time \cite{BBR}. Although the binary might stall at larger
separations due to the emptying of the loss cone of stars that could
extract angular momentum from its orbit (the so-called {\it final parsec
problem}), there are many plausible processes that would refill this loss
cone \cite{MM} or enable gas to grind the orbit down to the
gravitational-radiation dominated regime \cite{Escala}.  The end result is
likely to be a binary BH system that continues to tighten on its own due to
the emission of gravitational radiation, surrounded by an envelope of gas
\footnote{If the galaxy merger that produced the binary was devoid of cold
gas, it is possible for the binary to stall until the next galaxy merger
event. If a third BH joins the system, a slingshot ejection of the lightest
BH in the hierarchical triple BH system would occur \cite{Hoffman}. It is
possible to distinguish the gravitational-wave recoil case from the
slingshot ejection case by the fact that there should be a BH binary left
behind with its own accretion activity in the nucleus of the merger galaxy
for the slingshot case. If observations show lack of nuclear accretion
activity (in X-rays, for example) despite the existence of ample nuclear
gas, then the gravitational-wave recoil would offer a more plausible origin
for the ejection of an off-center quasar.}. The gas will likely settle to a
cold disk around the BH binary, since its cooling time is short. The disk
may include some gas that was originally associated with either BH
individually but joined into a common envelope once the binary separation
was sufficiently reduced.

Recent hydrodynamic simulations of a common co-rotating disk of gas around
a binary system of BHs of comparable mass \cite{Mac,Hay} indicate that a
hole tends to open across a region of a radius equal to twice the binary
semi-major axis $a$.  The clearing of the hole is similar to the opening of
a gap by a massive planet in a gaseous disk around a star \cite{Lubow}. The
torque exerted by the binary BH on the outer edge of the hole, pushes gas
elements outwards at that location \cite{Mac,Murray}.  The corresponding
transfer of angular momentum promotes the binary coalescence process
\cite{Arm,Mac}.  As the binary separation shrinks, the hole is expected to
gradually close from the outside due to the viscous transport of angular
momentum by the gas.  The viscous timescale of the disk can be expressed
\cite{Goodman} in terms of the $\alpha$--parameter \cite{Sh},
\begin{equation}
t_{\rm visc}(r)= 4.1\times 10^4~
\alpha_{-1}^{-0.8}\eta^{0.4}M_7^{1.2}r_3^{1.4}~{\rm yr},
\label{t_visc} 
\end{equation}
where $M_7$ is the total BH binary mass in units of $10^7M_\odot$, $r_3$ is
the radius $r$ relative to the binary center of mass in units of $10^3$
Schwarzschild radii of $M$ ($=2GM/c^2$), $\alpha_{-1}$ is the viscosity
parameter scaled to a characteristic value of 0.1, the opacity is assumed
to be dominated by Thomson scattering for a primordial gas composition, and
$\eta=(\epsilon/0.1)/(L/L_E)$ with $\epsilon$ being the radiative
efficiency and $L$ being the disk luminosity in units of the Eddington
limit $L_E=1.4\times 10^{45}M_7~{\rm erg~s^{-1}}$ if the $\alpha$--disk
were to extend down to the last stable orbit around a single BH.  For
comparison, the time it takes a binary on a circular orbit to coalesce due
to the emission of gravitational radiation is \cite{Teuk}
\begin{equation}
t_{\rm gw}=2\times 10^{6}~\left({M\over 4\mu}\right) a_3^4 M_7
~{\rm yr},
\label{t_gw}
\end{equation}
where $\mu=(M_1M_2/M)$ is the reduced mass, and $M=(M_1+M_2)$ is the total
mass of the binary with BH masses $M_1$ and $M_2$. As before, a subscript
$(...)_3$ denotes a lengthscale in units of $10^3$ Schwarzschild radii of
$M$.  Hereafter, we focus our attention on the case where the BH masses are
comparable ($M/\mu \sim 4$) since the recoil speed is small when one of
the binary members is much lighter than the other \cite{Schnitt}.

There is a minimum radius to the cavity around the binary, $R_{\rm in}$,
below which the decay time of the binary orbit will be much shorter than
the viscous time required to refill the cavity \cite{Mac,MP}. We obtain
this radius by requiring $t_{\rm gw}<t_{\rm visc}$ and substituting $r\sim
2a$ for the radius of the hole around the BH binary \cite{Mac}.  This
gives,
\begin{equation}
R_{\rm in,3}\approx  0.65 M_7^{0.077}\alpha_{-1}^{-0.31} \eta^{0.15}.
\label{R_in}
\end{equation}

As the binary orbit continues to tighten with $a \la 0.5 R_{\rm in}$, the
disk inner radius does not have sufficient time to close much farther
before the binary BH coalesces. The ejected BH would therefore carry an
initially punctured disk with an inner cavity radius of $R_{\rm in}$ and an
outer radius of $R_{\rm out} \approx {GM/v_{\rm ej}^2}$ or equivalently
\footnote{The binding energy per unit mass of a gas element with an initial
circular velocity ${\vec{v}}_{\rm c}$ around the BH is $-{1\over
2}\vert{\vec{v_{\rm c}}}\vert^2$.  The gas element will remain bound after
an impulsive change of $-{\vec{v}}_{\rm ej}$ in its velocity relative to
the BH, as long as $\vert -\vec{v}_{\rm ej}+\vec{v}_{\rm c}\vert < \sqrt{2}
\vert{\vec{v_c}}\vert$.  Averaging over possible orientations, we
approximate the outer radius of the captured disk as $\sim GM/v_{\rm
ej}^2$.},
\begin{equation}
R_{\rm out,3}\approx 45 v_8^{-2} .
\label{R_out}
\end{equation}
We therefore find that a bound disk could survive (i.e. $R_{\rm out}>R_{\rm
in}$) for the relevant regime of recoil speeds \footnote{Note that even if
the disk fragments to stars in its outer part due to the Toomre-$Q$
stability parameter dropping below unity \cite{Goodman1, Goodman}, the
stars interior to $R_{\rm out}$ would be carried together with the
remaining gas to accompany the ejected BH. The energy generated by stars or
supernova explosions could supplement the heat produced by MHD viscosity in
the disk and keep $Q\ga 1$. We ignore these complications in this {\it
Letter}.}.  Well interior to $R_{\rm out}$, the orbital velocity of the gas
exceeds $v_{\rm ej}$ and the gas elements maintain the adiabatic invariants
of their orbit around the relatively ``slow-moving'' BH.  This would hold
even if the BH is kicked in the plane of the disk. The gas outside $R_{\rm
out}$ is unable to respond sufficiently quickly to the ejection of the BH
and so it is left behind.  After the BH remnant is ejected, the captured
disk relaxes to a viscous equilibrium state and fills its central cavity on
a timescale $t_{\rm visc}(R_{\rm in})$, which is much shorter than its
global accretion lifetime $t_{\rm visc}(R_{\rm out})$. Magnetohydrodynamic
(MHD) simulations of an initially toroidal gas distribution demonstrate
this behaviour (see, e.g. Ref. \cite{Hawley} and references therein).

The total mass of the $\alpha$--disk \cite{Goodman} around the ejected BH
is,
\begin{equation}
M_{\rm disk}\approx 1.9\times 10^6~\alpha_{-1}^{-0.8} \eta^{-0.6}
M_7^{2.2} v_8^{-2.8}~M_\odot.
\label{M_disk}
\end{equation}
The above expression is valid only in the regime where $M_{\rm disk}\ll M$
since we ignored the mass of the disk (as well as its self-gravity) in the
overall momentum balance of the BH$+$disk system.  The condition that the
BH would not carry a disk mass in excess of $M$ implies
\begin{equation}
R_{\rm out,3}< R_{\rm max,3}\equiv 1.5\times 10^2
\alpha_{-1}^{0.57}\eta^{-0.43} M_7^{-0.86}.
\label{lim}
\end{equation}
For the characteristic surface density and pressure of an $\alpha$-disk
\cite{Goodman}, we find that the ram-pressure of the ISM through which it
passes (as well as the amount of ISM mass intercepted) can be ignored.

\paragraph{Observable Signatures.}
A gas-rich merger of two galaxies naturally creates an environment that
includes a binary BH system surrounded by gas near the center of the merger
product \cite{Hernquist2,Escala}. As gas accumulates outside the binary, it
is likely to form a punctured disk that would survive through the binary
coalescence process.

The lifetime of a disk with an outer radius $R_{\rm out}$ given by
Eq. (\ref{R_out}) can be derived \footnote{In the regime where $R_{\rm
max}$ in Eq. (\ref{lim}) is smaller than $R_{\rm out,3}=45v_8^{-2}$ in
Eq. (\ref{R_out}), the values of $M_{\rm disk}$ in Eq. (\ref{M_disk}) and
$t_{\rm disk}$ in Eq. (\ref{life}) are both reduced by a factor $\ga
(R_{\rm max,3}/45v_8^{-2})^{1.4}$.} from Eq. (\ref{t_visc}),
\begin{equation}
t_{\rm disk}\approx 8.4\times 10^6~\alpha_{-1}^{-0.8}\eta^{0.4}M_7^{1.2}
v_8^{-2.8}~{\rm yr} .
\label{life}
\end{equation}
During the disk lifetime $t_{\rm disk}$, the
ejected BH could traverse a distance of
\begin{equation}
d\approx v_{\rm ej} t_{\rm disk}\approx 8.6~
\alpha_{-1}^{-0.8}\eta^{0.4}M_7^{1.2} v_8^{-1.8}~{\rm kpc} ,
\label{offset}
\end{equation}
and appear as an off-center quasar. An offset of $d\sim 10$ kpc can be
resolved, as it corresponds to angular scales of order an arcsecond at
cosmological distances.  A noticeable displacement might exist even for BHs
which remain bound to their host galaxy halo.  There has already been a
claim for a detection of a displaced quasar \cite{Magain}, which has been
disputed by subsequent analysis \cite{Merritt}.  An undisputed example for
the preceding phase of a compact BH binary was reported recently in a
different system \cite{Rodriguez}.  

Interestingly, the offset in Eq. (\ref{offset}) increases as the ejection
speed decreases.  This scaling originates from the steep inverse dependence
of the disk size on the ejection speed.  However, it is only valid as long
as $v_{\rm ej}\gg 200~{\rm km~s^{-1}}$, because the accretion disk is not
expected to follow the ejected BH from radii where the gas orbital speed is
lower than $\sim 200~{\rm km~s^{-1}}$ at which the tidal force from the
galaxy is important.  For low ejection speeds, $v_{\rm ej}\la \sigma$, the
BH would only carry the fraction of the disk interior to $R_{\rm out}\sim
\min\{(GM/\sigma^2),R_{\rm max}\}$, where $\sigma$ is the velocity
dispersion of the host galactic spheroid. Expressing
$\sigma=200\sigma_{200}~{\rm km~s^{-1}}$, the value of
$(GM/\sigma^2)=1.1\times 10^3 \times (10^3\times 2GM/c^2)\sigma_{200}^{-2}$
is typically larger than $R_{\rm max}$ for $M_7\ga 1$ in Eq.  (\ref{lim})
and hence is of secondary importance.  At low ejection speeds, the
deceleration of the BH by the external gravitational potential of the
galaxy and by dynamical friction on the background stars and gas, would
limit the displacement of the slow-moving BH and tend to bring it back to
the center of its host galaxy within a few dynamical times
\cite{Quataert,Hoffman}.

For a high ejection speed, $v_8\ga 1$, the displacement of the ejected BH
relative to the host galaxy would also be noticeable in velocity (redshift)
space.  The velocity offset could be detected if the material bound to the
ejected BH (including gas and possibly some captured stars) is capable of
producing emission lines.

The typical lifetime of quasars is estimated \cite{Martini} to be
comparable to the value of $t_{\rm disk}$ in Eq. (\ref{life}).  However,
for equal-mass binaries with random spins (uniformly distributed between
dimensionless values of 0 and 0.9 and of random orientations) only $\sim
15\pm 6\%$ of all mergers are expected to result in an ejection speed
$v_{8}>0.5$ and only $\sim 3\pm 2\%$ in $v_{8}>1$ \cite{Schnitt}, above the
escape speed of a massive galaxy. Hence, we expect only a small fraction of
all quasars to be associated with an unbound BH that escapes the
galaxy. Indeed, the fact that almost all nearby galaxies possess a nuclear
BH \cite{Tremaine} even though some formed out of gas-poor mergers (from
which a new BH could not have been created to replace an expelled BH)
implies that unbinding kicks are rare.  Similarly to most bright quasars,
the luminosity of a kicked BH could be close to the Eddington limit, except
that its fuel reservoir is short-lived and cannot be replenished unless
$v_{\rm ej}\la \sigma$ and the BH settles back to the center of the galaxy.

The motion of the BH and its displacement to regions of low gas density
would reduce the growth rate of its mass relative to the case where the
remnant BH stays at the center of its host galaxy. By the time an ejected
BH is driven back to the center by dynamical friction, a substantial
fraction of the cold gas there may already be converted into stars or
expelled in a galactic wind.

The feedback on the host galaxy from the radiation or wind produced by
the accreting disk around the ejected BH would obviously depend on the
ejection speed. Previous treatments of BH feedback in galaxy mergers
\cite{Hernquist2} ignored the gravitational-wave recoil and assumed that
the BH remnant stays fixed at the center of mass of its host galaxy. The
energy/momentum feedback was therefore confined to the innermost (and hence
densest) resolution element of gas. However, the actual probability
distribution of ejection speeds \cite{Schnitt} allows for substantial BH
displacements in many cases, for which the energy/momentum output from the
BH would interact differently with the surrounding gas.  If the ejected BH
carries an accretion disk as discussed in this paper, its feedback on the
host galaxy would be different than previously calculated.  The distributed
energy/momentum release of a displaced BH would leave a trail of evidence
that traces its origin to the center of the galaxy through the mark of
its feedback along its path. For example, this mark could be traced by
observing enhanced H$\alpha$ line emission due to the ionization trail
imprinted by the BH on the diffuse ISM along its trajectory.

The basic conclusions of this {\it Letter} would be altered if gas is
expelled from the vicinity of the BH binary by a powerful galactic wind,
driven by supernovae or quasar activity prior to the binary merger event. A
burst of star formation activity is generically expected to accompany the
merger process \cite{Hernquist2} and could affect the assembly of cold gas
into the circumbinary accretion disk that was postulated in this {\it
Letter}.  It is also possible that low-level accretion across the gap
surrounding the BH binary, would consume the gas in the initial disk and
weaken the late accretion episode considered here.

In the future, it would be useful to improve upon the approximate treatment
presented here by simulating the dynamical response of an MHD disk to the
final stage of the BH binary coalescence process and the subsequent recoil
of the BH remnant. Such a calculation can be done with existing codes
\cite{Mac,Hay} that were used so far to address other aspects of the binary
coalescence phenomenon. Cosmological hydrodynamics codes \cite{Hernquist2}
could also be used to simulate the feedback trail that an ejected BH
imprints on the ISM of its host galaxy, which is reminiscent of the track
of an elementary particle in a bubble chamber.

\bigskip
\bigskip
\bigskip
\bigskip

\paragraph*{Acknowledgments}

I thank L. Hoffman, M. Holman, R. O'Leary and G. Rybicki for helpful
comments on the manuscript.

\end{document}